\documentclass[aps,prl,twocolumn,superscriptaddress,showpacs,nofootinbib]{revtex4-1}
\usepackage{graphicx}
\usepackage{color}

\begin{document}

% Use the \preprint command to place your local institutional report
% number in the upper righthand corner of the title page in preprint mode.
% Multiple \preprint commands are allowed.
% Use the 'preprintnumbers' class option to override journal defaults
% to display numbers if necessary
%\preprint{}

%Title of paper
\title{
%Dipole Polarizability of $^{120}$Sn via Complete Measurement of Electric Dipole Responce
Test of the Brink-Axel Hypothesis for the Pygmy Dipole Resonance
}

% repeat the \author .. \affiliation  etc. as needed
% \email, \thanks, \homepage, \altaffiliation all apply to the current
% author. Explanatory text should go in the []'s, actual e-mail
% address or url should go in the {}'s for \email and \homepage.
% Please use the appropriate macro foreach each type of information

% \affiliation command applies to all authors since the last
% \affiliation command. The \affiliation command should follow the
% other information
% \affiliation can be followed by \email, \homepage, \thanks as well.

\newcommand{\RCNP}{Research Center for Nuclear Physics, Osaka University, Ibaraki, Osaka 567-0047, Japan}
\newcommand{\Wits}{School of Physics, University of the Witwatersrand, Johannesburg 2050, South Africa}
\newcommand{\Kyushu}{Department of Physics, Kyushu University, Fukuoka 812-8581 ,Japan}
\newcommand{\iThemba}{iThemba LABS, Somerset West 7129, South Africa}
\newcommand{\Osaka}{Department of Physics, Osaka University, Toyonaka, Osaka 560-0043, Japan}
\newcommand{\CYRIC}{Cyclotron and Radioisotope Center, Tohoku University, Sendai,  980-8578, Japan}
\newcommand{\CNS}{Center for Nuclear Study, University of Tokyo, Bunkyo,  Tokyo 113-0033, Japan}
\newcommand{\TUDarmstadt}{Institut f\"{u}r Kernphysik, Technische Universit\"{a}t Darmstadt, D-64289 Darmstadt,
Germany}
\newcommand{\Valencia}{Instituto de Fisica Corpuscular, CSIC-Univ. de Valencia, E-46071 Valencia, Spain}
\newcommand{\Gent}{Vakgroep Subatomaire en Stralingsfysica, Univ. Gent, B-9000 Gent, Belgium}
\newcommand{\MSU}{NSCL, Michigan State Univ., MI 48824, USA}
\newcommand{\Kyoto}{Department of Physics, Kyoto University, Kyoto 606-8502, Japan}
\newcommand{\Niigata}{Department of Physics, Niigata University, Niigata 950-2102, Japan}
\newcommand{\RIKEN}{RIKEN Nishina Center, Wako, Saitama 351-0198, Japan}
\newcommand{\HIMAC}{National Institute of Radiological Sciences, Chiba 263-8555, Japan}
\newcommand{\KVI}{Kernfysisch Versneller Instituut, University of Groningen, Zernikelaan 25, NL-9747 AA Groningen,
The Netherlands}
\newcommand{\TexasAM}{Department of Physics and Astronomy, Texas A\&M University-Commerce, Commerce, Texas 75429,
USA}
\newcommand{\GSI}{GSI Helmholtzzentrum f\"{u}r Schwerionenforschung, 64291 Darmstadt, Germany}
\newcommand{\INST}{Institute for Nuclear Science and Technology, 179 Hoang Quoc Viet, Hanoi, Vietnam}
\newcommand{\UEN}{Institut f\"ur Theoretische Physik, Universit\"at Erlangen, D-91054 Erlangen, Germany}
\newcommand{\OsakaU}{Department of Physics, Osaka University, Toyonaka, Osaka, 560-0043, Japan}
\newcommand{\Istanbul}{Physics Department, Faculty of Science, Istanbul University, 34459 Vezneciler, Istanbul,
Turkey}
\newcommand{\Munstar}{Institut f\"{u}r Kernphysik, Westf\"{a}lische Wilhelms-Universt\"{a}t M\"{u}nster, D-48149
M\"{u}nster, Germany}
\newcommand{\Daejeon}{Rare Isotope Project, Institute for Basic Science, 70, Yuseong-daero, 1689-gil,Yuseong-gu, Daejeon, Korea}
\newcommand{\Myazaki}{Faculty of Engineering, Miyazaki University, Miyazaki 889-2192, Japan}
\newcommand{\Tohoku}{Department of Physics, Tohoku University, Sendai, Miyagi 980-8578, Japan}
\newcommand{\TWMU}{Tokyo Women's Medical University, Tokyo 162-8666, Japan}

\author{D.~Martin}\affiliation{\TUDarmstadt}
\author{P.~von Neumann-Cosel}\email[E-mail: ]{vnc@ikp.tu-darmstadt.de}\affiliation{\TUDarmstadt}
\author{A.~Tamii}\affiliation{\RCNP}
\author{N.~Aoi}\affiliation{\RCNP}
\author{S.~Bassauer}\affiliation{\TUDarmstadt}
\author{C.~A.~Bertulani}\affiliation{\TexasAM}
%\author{B.~Bilgier}\affiliation{\Istanbul}
\author{J.~Carter}\affiliation{\Wits}
\author{L.~Donaldson}\affiliation{\Wits}
\author{H.~Fujita}\affiliation{\RCNP}
\author{Y.~Fujita}\affiliation{\RCNP}
%\author{E. Ganio\v{g}lu}\affiliation{\Istanbul}
\author{T. Hashimoto}\affiliation{\RCNP} 
\author{K.~Hatanaka}\affiliation{\RCNP}
\author{T.~Ito}\affiliation{\RCNP}
%\author{C.~Iwamoto}\affiliation{\RCNP}
%\author{T.~Kawabata}\affiliation{\Kyoto}
%\author{C.~Kozer}\affiliation{\Istanbul}
\author{A.~Krugmann}\affiliation{\TUDarmstadt}
%\author{A.~M.~Krumbholz}\affiliation{\TUDarmstadt}
%\author{J.~Lee}\affiliation{\RIKEN}
\author{B.~Liu}\affiliation{\RCNP}
\author{Y.~Maeda}\affiliation{\Myazaki}
%\author{H.~Matsubara}\affiliation{\HIMAC}\affiliation{\TWMU}
\author{K.~Miki}\affiliation{\RCNP}
%\author{M.~Nagashima}\affiliation{\Niigata}
\author{R.~Neveling}\affiliation{\iThemba}
%\author{H.~J.~Ong}\affiliation{\RCNP}
\author{N.~Pietralla}\affiliation{\TUDarmstadt}
\author{I.~Poltoratska}\affiliation{\TUDarmstadt}
\author{V.~Yu.~Ponomarev}\affiliation{\TUDarmstadt}
\author{A.~Richter}\affiliation{\TUDarmstadt}
%\author{H.~Sakaguchi}\affiliation{\RCNP}
\author{T.~Shima}\affiliation{\RCNP}
%\author{Y.~Shimbara}\affiliation{\Niigata}
%\author{F.~D.~Smit}\affiliation{\iThemba}
%\author{G.~S\"{u}soy}\affiliation{\Istanbul}
%\author{T.~Suzuki}\affiliation{\RCNP}
%\author{M.~Wiedeking}\affiliation{\iThemba}
\author{T.~Yamamoto}\affiliation{\RCNP}
%\author{M.~Yosoi}\affiliation{\RCNP}
%\author{J.~Zenihiro}\affiliation{\RIKEN}
\author{M.~Zweidinger}\affiliation{\TUDarmstadt}

%Collaboration name if desired (requires use of superscriptaddress
%option in \documentclass). \noaffiliation is required (may also be
%used with the \author command).
%\collaboration can be followed by \email, \homepage, \thanks as well.
%\collaboration{}
%\noaffiliation

\date{\today}

\begin{abstract}
The gamma strength function (GSF) and level density (LD) of $1^-$ states in $^{96}$Mo have been extracted from a high-resolution study of the $(\vec{p},\vec{p}^\prime)$ reaction at 295 MeV and extreme forward angles. 
By comparison with compound nucleus $\gamma$ decay experiments, this allows a test of the generalized Brink-Axel (BA) hypothesis in the energy region of the Pygmy Dipole Resonance (PDR).
The BA hypothesis is commonly assumed in astrophysical reaction network calculations and states that the GSF in nuclei is independent of the structure of initial and final state.
The present results validiate the BA hypothesis for $^{96}$Mo and provide independent confirmation of the methods used to separate GSF and LD in $\gamma$ decay experiments.      
\end{abstract}

% insert suggested PACS numbers in braces on next line
%\pacs{23.20.Lv, 24.30.Gd, 25.70.De, 27.60.+j}
% insert suggested keywords - APS authors don't need to do this
%\keywords{}

%\maketitle must follow title, authors, abstract, \pacs, and \keywords
\maketitle

% body of paper here - Use proper section commands
% References should be done using the \cite, \ref, and \label commands
%\section{Intoroduction}
% Put \label in argument of \section for cross-referencing
%\section{\label{}}
%\subsection{}
%\subsubsection{}

%\section{Introduction}

{\it Introduction.}--The gamma strength function describes the average $\gamma$ decay behavior of a nucleus.
Their knowledge is required for applications of statistical nuclear theory in astrophysics \cite{arn07}, reactor design \cite{cha11}, and waste transmutation \cite{sal11}.
Although all electromagnetic mulipoles contribute, the GSF is dominated by the E1 radiation with smaller contributions from M1 strength.
Above particle threshold it is governed by the IsoVector Giant Dipole Resonance (IVGDR) but at lower excitation energies the situation is complex:
In nuclei with neutron excess one observes the formation of the Pygmy Dipole Resonance (PDR) \cite{sav13} sitting on  the low-energy tail of the IVGDR.
The impact of low-energy E1 strength on astrophysical reaction rates and the resulting abundances in the $r$ process have been discussed e.g.\ in Refs. \cite{gor04,lit09,dao12}.

Most applications imply an environment of finite temperature, notably in stellar scenarios \cite{wie12}, and thus reactions on initially excited states become relevant.
Their contributions to the reaction rates are usually estimated applying the generalized Brink-Axel (BA) hypothesis \cite{bri55,axe62}, which states that the GSF is independent of the properties of the initial and final states and thus should be the same in $\gamma$ emission and absorption experiments.
Although historically formulated for the IVGDR, where it seems to hold approximately for not too high temperatures \cite{bbb98}, this is nowadays a commonly used assumption to calculate the low-energy E1 and M1 strength functions. 
Recent theoretical studies \cite{joh15,hun17} put that into question demonstrating that the strength functions of collective modes built on excited states do show an energy dependence.
However, numerical results for E1 strength functions obtained in Ref.~\cite{joh15}  showed an approximate constancy as a function of excitation energy consistent with the BA hypothesis. 

Recent work utilizing compound nucleus $\gamma$ decay with the so-called Oslo method \cite{sch00} has demonstrated independence of the GSF from excitation energies and spins of initial and final states in a given nucleus in accordance with the BA hypothesis once the level densities are sufficiently high to suppress large intensity fluctuations \cite{gut16}.
However, there are a number of experimental results which indicate violations of the BA hypothesis in the low-energy region.
For example, the GSF in heavy deformed nuclei at excitation energies of $2 -3$ MeV is dominated by the orbital M1 scissors mode \cite{boh84} and potentially large differences in $B$(M1) strengths are observed between $\gamma$ emission  and absorption  experiments \cite{hey10,gut12,ang16}. 
At very low energies ($< 2$ MeV) an increase of GSFs is observed in Oslo-type experiments \cite{voi04,lar14}, which for even-even nuclei cannot have a counterpart in ground state absorption experiments because of the pairing gap.           

For the low-energy E1 strength in the region of the PDR, the question is far from clear when comparing results from the Oslo method with photoabsorption data.
Below particle thresholds most information on the GSF stems from nuclear resonance fluorescence (NRF) experiments.
A striking example of disagreement is the GSF of $^{96}$Mo, where the results from the Oslo method \cite{gut05,lar10} and from NRF \cite{rus09} differ by factors $2 -3$ at excitation energies between 4 and 7 MeV.
A problem with the NRF method are the experimentally unobserved braching ratios to excited states. 
While many older analyses of NRF data assume that these are negligible, in Ref.~\cite{rus09} they are included by a Hauser-Feshbach calculation assuming statistical decay.
The resulting corrections are sizable, reaching a factor of five close to the neutron threshold.
On the other hand, there are indications of non-statistical decay behavior of the PDR from recent measurements \cite{rom15,loe16}.
Violation of the BA hypothesis was also claimed in a simultaneous study of the $(\gamma,\gamma^\prime)$ reaction and average ground state branching ratios \cite{ang12} in $^{142}$Nd (see, however, Ref.~\cite{ang15}).   
Clearly, information on the GSF -- in particular in the energy region of the PDR -- from an independent experiment is called for.

A new method for the measurement of complete E1 strength distributions -- and thereby the E1 part of the GSFs -- in nuclei from about 5 to 25 MeV has been developed using relativistic Coulomb excitation in polarized inelastic proton scattering at energies of a few hundred MeV and  scattering angles close to $0^\circ$ \cite{tam11,pol12,kru15,has15,bir17}.
These data allow the dipole polarizability to be determined which provides important constraints on the neutron skin of nuclei and the poorly known parameters of the symmetry energy \cite{epj50}.
They alsopermit extraction of the M1 part of the GSF due to spinflip excitations \cite{bir16}, which energetically overlaps with the PDR strength.
Furthermore, when performed with good energy resolution, the level density can be extracted in the excitation region of the IVGDR from the giant resonance fine structure independent of the GSF \cite{pol14}.
This allows an important test of the model-dependent decomposition of LD and GSF in the Oslo method \cite{sch00} . 

Such a test has been performed for the case of $^{208}$Pb \cite{bas16} and good agreement of LDs deduced from the Oslo method and the $(p,p^\prime)$ data was found.
However, because of the extremely low LD of a doubly magic nucleus and the corresponding strong intensity fluctations in a ground-state absorption experiment,  the comparison of the GSFs in the PDR energy region remained inconclusive. 
Here, a study of an open-shell nucleus is reported, where the LD should be sufficiently high to a comparison of averaged quantities from the $(p,p^\prime)$ experiment.   
The case of $^{96}$Mo was selected for a critical examination of the above-discussed apparent violation of the BA hypothesis in the low-energy regime suggested by the NRF data \cite{rus09}. 

{\it Experiment.}--The $^{96}$Mo($\vec{p},\vec{p}\,^\prime$) reaction was studied at RCNP, Osaka, Japan. 
Details of the experimental techniques can be found in Ref.~\cite{tam09}.
A proton beam of 295 MeV with intensities of about  2 nA  at $0^\circ$ up to 6 nA at larger spectrometer angles and with an average polarization  $P_0 \simeq 0.67$ impinged on a $^{96}$Mo foil isotopically enriched to 96.7\% with an areal density of 3 mg/cm$^2$.
Data were taken with the Grand Raiden spectrometer \cite{fuj99} placed at $0^\circ$ covering an angular acceptance of $\pm 2.6^\circ$ and excitation energies $E_x \simeq 5 - 23$ MeV.
The energy resolution varied between 25  and 40~keV (full width at half maximum, FWHM).
Normally ($N$) and longitudinally ($L$) polarized beams were used to measure the polarization transfer coefficients \cite{ohl79} $D_{NN'}$ and $D_{LL'}$, respectively.
Additional data with unpolarized protons were taken for angles up to $6^\circ$.
%Utilizing dispersion matching techniques, a high energy resolution $\Delta E \simeq 25$ keV (full width at half maximum) could be achieved.
%
\begin{figure}[t]
\includegraphics[width=8cm]{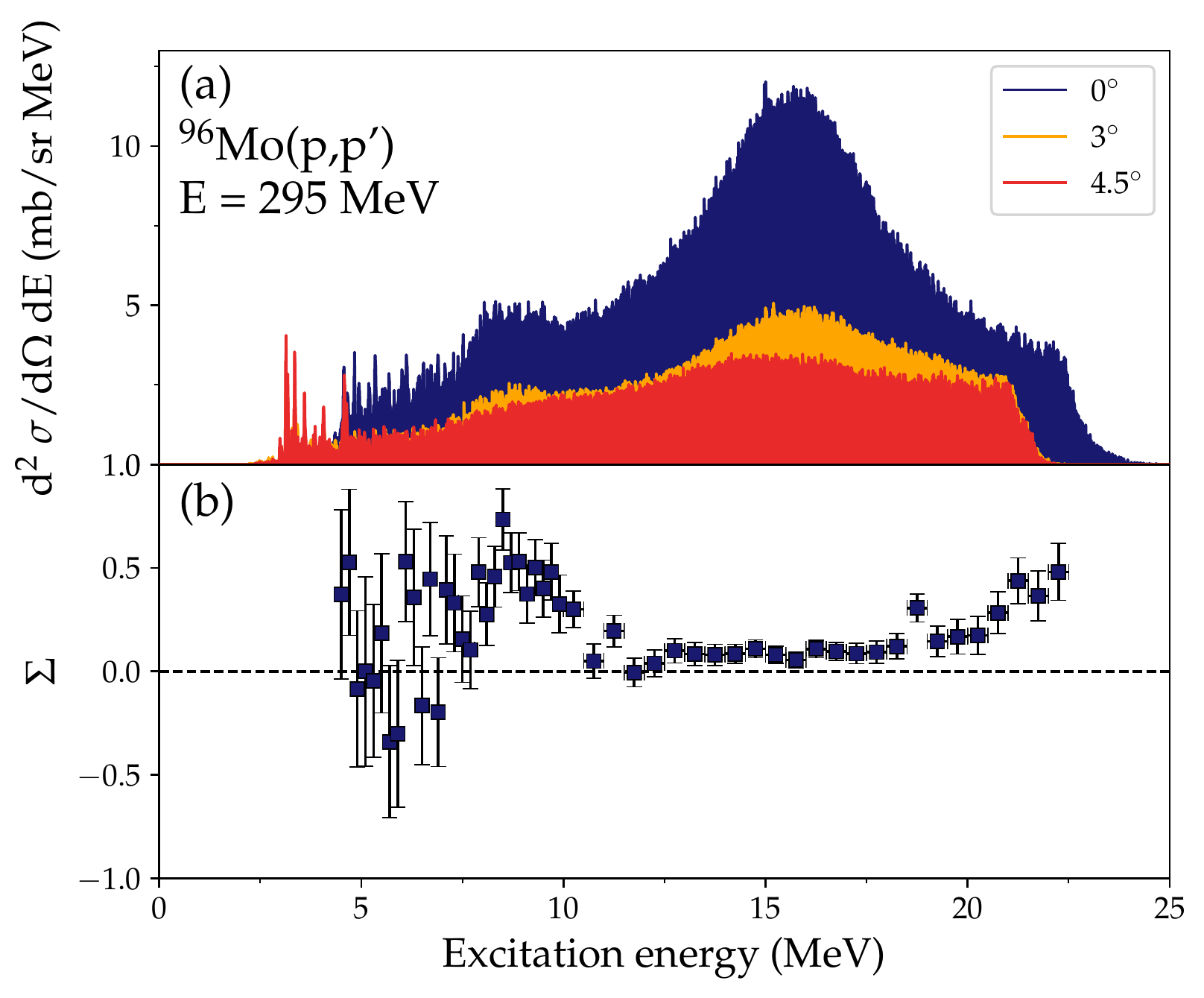}
\includegraphics[width=8.6cm]{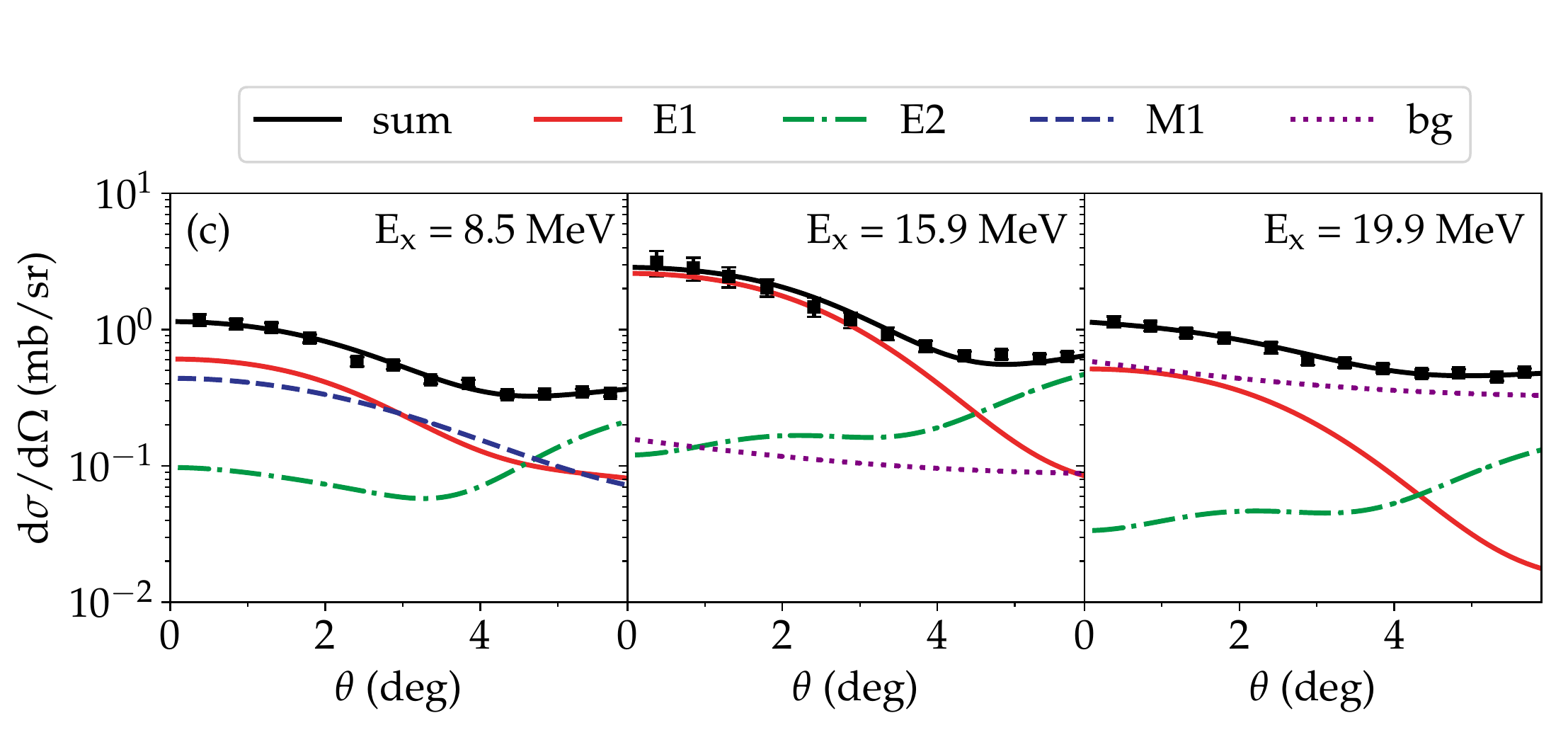}
\caption{\label{fig:spec}
(Color online) (a) Spectra of the $^{96}$Mo($\vec{p},\vec{p}\,^\prime$) reaction at $E_p = 295$ MeV with the spectrometer placed at $0^\circ$ (blue), $3^\circ$ (yellow), and $4.5^\circ$ (red). (b) Total spin transfer $\Sigma$, Eq.~(\ref{eq:spintransfer}).  (c) Examples of the MDA for selected energy bins.}
\end{figure}

Figure~\ref{fig:spec}(a) displays the spectra taken at spectrometer angles $0^\circ$, $3^\circ$, and $4.5^\circ$. 
Besides discrete transitions at low excitation energies, a resonance-like structure around 8 MeV and the prominent IVGDR centered at $E_x \approx 16$ MeV are observed.
The cross sections show a strong forward-angle peaking indicating the dominance of Coulomb excitation. 
The total spin transfer 
\begin{equation}
  \Sigma = \frac{3-(2D_{NN'}+D_{LL'})}{4}
  \label{eq:spintransfer}
\end{equation}
shown in Fig.~\ref{fig:spec}(b)  can be extracted from the measured polarization transfer observables.
It takes a value of one for spinflip ($\Delta S = 1$) and zero for non-spinflip ($\Delta S = 0$) transitions at $0^\circ$ \cite{suz00}, interpreted as M1 and E1 cross sections parts, respectively.
Values in between 0 and 1 result from the summation over finite energy bins (200 keV up to an excitation energy of 10 MeV and 500 keV above). 
The polarization transfer analysis (PTA) reveals the presence of a few spinflip transitions between 5 and 7.5 MeV and a concentration of spinflip strength in the energy region $ 7.5 - 10$ MeV identified as the spin-M1 resonance in $^{96}$Mo. 
Cross sections above 10 MeV are dominantly of $\Delta S = 0$ character as expected for Coulomb excitation.
These findings are consistent with the results in $^{208}$Pb \cite{tam11} and $^{120}$Sn \cite{has15}.
% Shortened 24.5.11
The $\Delta S = 1$ strength observed at high $E_{\rm x}$  can be understood to arise from quasi-free scattering \cite{bak97}.

Alternatively, a multipole decomposition analysis (MDA) was performed for angular distributions of the cross sections in the PDR and GDR regions.
For this purpose, angular cuts were applied to the spectra of Fig.~\ref{fig:spec}(a) providing 4 data points each.   
The MDA followed closely the approach described in Refs.~\cite{tam11,pol12}.
For the nuclear background the empirical parametrization found for $^{208}$Pb \cite{pol11} was adopted assuming the same momentum transfer dependence.
Figure \ref{fig:spec}(c) presents representative examples of the MDA for 200 keV excitation energy bins at different excitation energies.
They demonstrate that in the low-energy bump M1 contributions are sizable while E1 dominates in the energy region of the IVGDR. 
At even higher energies, the nuclear background becomes dominant.

\begin{figure}[t]
\includegraphics[width=8.6cm]{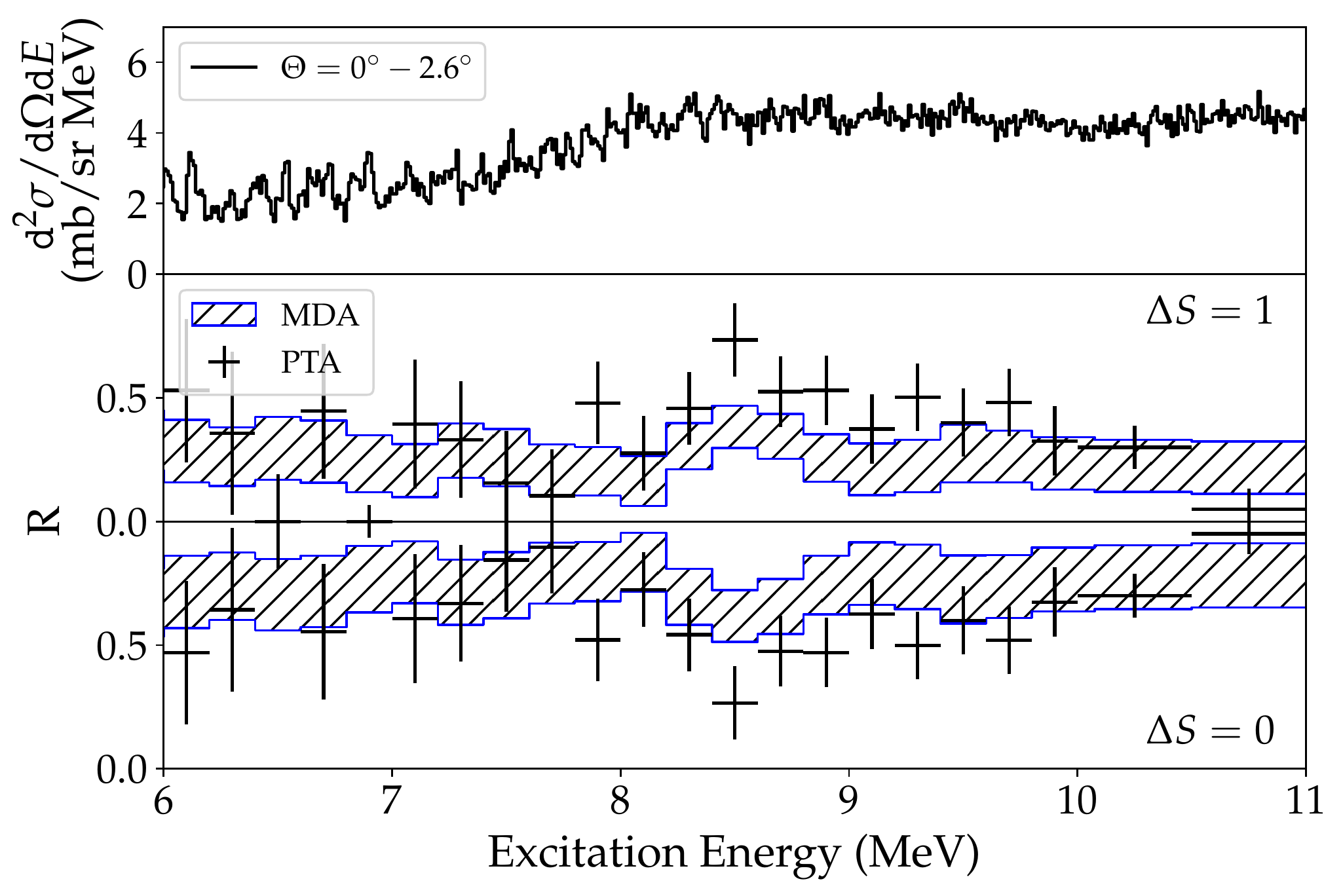}
\caption{\label{fig:compare}
(Color online) Relative yield $R$ of non-spinflip ($\Delta S = 0$) and spinflip ($\Delta S = 1$) cross section parts of the $^{96}$Mo($\vec{p},\vec{p}\,^\prime$) spectrum in the excitation energy region $6 - 11$ MeV based on the MDA and PTA, respectively. 
Agreement between the two independent methods is observed within error bars.
}
\end{figure}
The relative yield $R$  of spinflip and non-spinflip cross sections resulting from the MDA and PTA for $E_x \leq 11$~MeV is compared in Fig.~\ref{fig:compare}. 
The two completely independent decomposition methods lead to consistent results within error bars except for one energy bin around 8.5 MeV.
In the following, E1 and M1 cross sections averaged over both decomposition methods are considered for excitation energies up to 11 MeV.
At higher $E_{\rm x}$ only the MDA results are taken since the $\Delta S = 0$ part of the nuclear background, which cannot be distinguished in the PTA, becomes relevant.   

{\em Gamma Strength Function.}--The GSF for electric or magnetic transitions $X\in{\left\lbrace E,M\right\rbrace }$ with multipolarity $\lambda$ is related to the photoabsorption cross section $\left\langle\sigma_{abs}^{X\lambda}\right\rangle$ by
\begin{equation}
	\label{eqn:gsf}
	f^{X\lambda}(E, J) = \frac{2J_0+1}{(\pi\hbar c)^2 (2J+1)}
	\frac{\left\langle\sigma_{abs}^{X\lambda}\right\rangle}{E_\gamma^{2\lambda-1}} ,
\end{equation}
where  $E_\gamma$ denotes the $\gamma$ energy and $J, J_0$ are the spins of excited and ground state, respectively~\cite{cap09}.
For absorption experiments $E_{\rm x} = E_\gamma$.
The brackets $\langle \rangle$ indicate averaging over an energy interval.
In practice only E1 and M1 transitions provide sizable contributions to the total GSF. 
The Coulomb excitation cross sections representing the E1 part of the GSF were converted to equivalent photoabsorption cross sections using the virtual photon method \cite{ber88}.
The virtual photon spectrum exhibits a strong energy dependence, which amounts to a factor of ten for the energy region $ 6 - 20$ MeV covered in the present experiment.
It was calculated in an eikonal approach \cite{ber93} and integrated over the solid angle covered by the experiment. 
%The minimum impact parameter was estimated from the sum of the mean square radius of the proton and a matter radius of $^{96}$Mo given by $r = 1.2 A^{1/3}$.    
The M1 cross sections from Fig.~\ref{fig:compare} were converted to reduced transition strengths and the corresponding M1 photoabsoprtion cross sections with the approach described in Refs.~\cite{bir16,mat17}.

\begin{figure}[tbh]
\includegraphics[width=8.6cm]{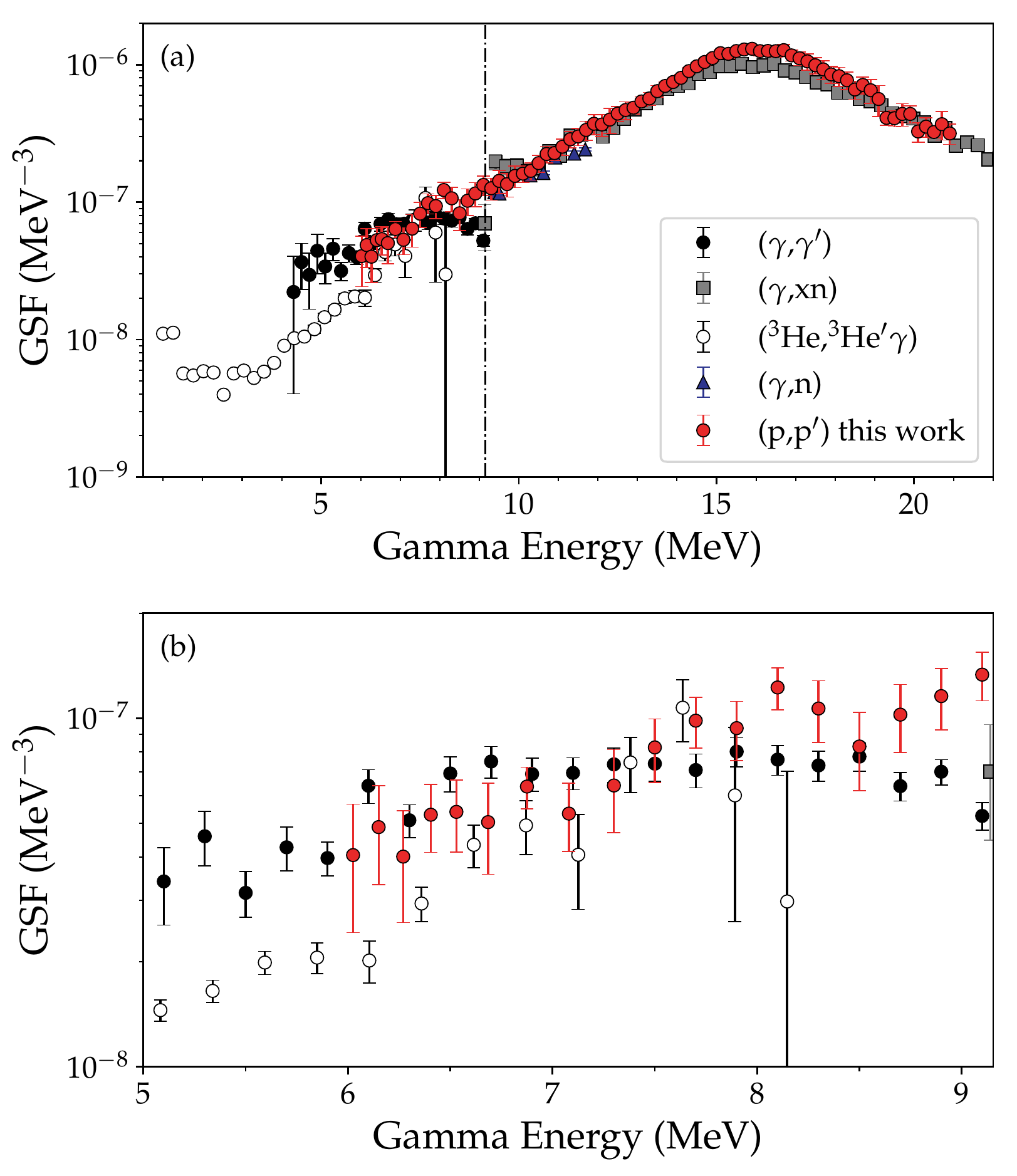}
\caption{\label{fig:gsf}
(Color online). 
(a) GSF of $^{96}$Mo from the present work (red circles) compared with ($^3{\rm He},^3{\rm He}^\prime \gamma$)  \cite{gut05,lar10} (open circles),  ($\gamma,xn$) \cite{bei74} (grey squares), ($\gamma,n$) \cite{uts13}, (blue upward triangles) and ($\gamma,\gamma^\prime$) \cite{rus09} data including a statistical model correction for unobserved branching ratios (black circles)
(b) Expanded range from 5 MeV to neutron threshold.   
}
\end{figure}
The sum approximating the total GSF in $^{96}$Mo is displayed in Fig.~\ref{fig:gsf}(a) as red circles for $E_{\gamma} = 6 - 20$ MeV. 
The error bars include statistical (dominating the PTA) and systematic (dominating the MDA) uncertainties 
The result is compared with ($^3$He,$^3$He$^\prime\gamma$)  \cite{gut05,lar10} (open circles),  ($\gamma,xn$) \cite{bei74} (grey squares), ($\gamma,n$) \cite{uts13} (blue upward triangles), and ($\gamma,\gamma^\prime$) data corrected for unobserved branching ratios \cite{rus09}  (black circles).
Above threshold, there is overall fair agreement with the data from Refs. ~\cite{bei74,uts13} except that the present experiment finds somewhat larger photoabsorption cross sections around the maximum of the IVGDR.

Below $S_n$, the GSF from the present work lies in between the Oslo and the $(\gamma,\gamma^\prime)$ experiment.
An expanded view of the GSF results between 5 MeV and the neutron threshold $S_n = 9.154$ MeV is displayed in Fig.~\ref{fig:gsf}(b).
The $(p,p^\prime$)  and ($^3$He,$^3$He$^\prime\gamma$) results agree within error bars except for the two lowest excitation energies analyzed in the present data.
However, these two data points suffer from limited statistics.
The ($\gamma,\gamma^\prime$) results \cite{rus09} agree in the $7 - 8$ MeV excitation energy region $7 - 8$ MeV but clearly underestimate the present results at higher $E_{\rm x}$.
At lower $E_{\rm x}$ they are systematically at the upper limit of the present results (and sometimes exceed it) and are significantly larger than the Oslo results.
The deviations from the present results may arise from the modeling of the large atomic background in the spectra and/or the specific choice of level densities for the simulation of the $\gamma$ decay cascades \cite{rus08}. 
     
\begin{figure}[tbh]
\includegraphics[width=8.6cm]{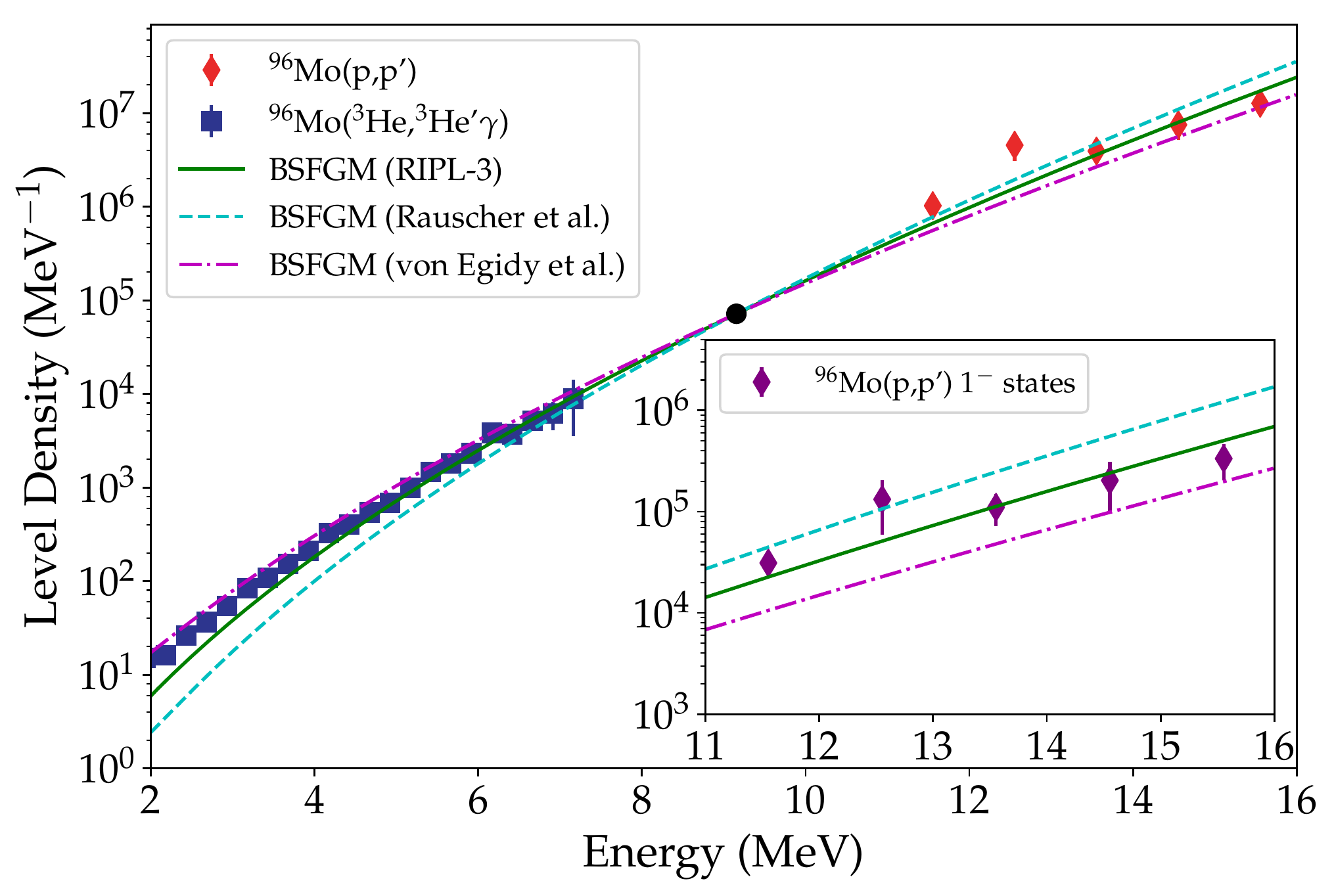}
\caption{\label{fig:ld}
(Color online). 
Total LD in $^{96}$Mo deduced from the fine structure of the $(p,p^\prime)$ data in the energy region of the IVGDR (red diamonds) compared with the results  from  the ($^3$He,$^3$He$^\prime\gamma$) Oslo experiment (blue squares) \cite{gut05,lar10}.
The black circle point stems from $s$-wave resonance neutron capture \cite{uts13}.  
BSFG models normalized to the value at $S_n$ are shown as green solid \cite{cap09}, cyan dashed \cite{rau97}, and purple dashed-dotted \cite{egi09} lines.
The inlet shows the LD of $1^-$ states in comparison with absolute predictions of the models.
}
\end{figure}

{\em Level Density.}--Since only the product of GSF and LD is measured by the Oslo method \cite{sch00}, it relies on external data for their decomposition. 
An independent check of the LD results for $^{96}$Mo is thus of high importance.
The good energy resolution of the present data permits an extraction of the LD of $J^\pi = 1^-$ states applying a fluctuation analysis to the fine structure of the IVGDR..
Details of the method can be found in Refs.~\cite{pol14,kal06,usm11}.
We note that the method is based on the assumptiion of a single class of excited states in the spectrum.
This presently limits the application to the energy region of the IVGDR while at lower excitation energies $1^-$ and $1^+$ states coexist, since the PDR and the spin M1 strength overlap.  
The LD of $J^\pi = 1^-$ states between 11 and 16 MeV is displayed in the inlet of Fig.~\ref{fig:ld} in comparison with three widely used systematic parametrizations \cite{cap09,rau97,egi09} of the phenomenological backshifted Fermi gas (BSFG) model  (see Table \ref{tab:ld}). 
The BSFG parameters deduced from the RIPL-3 data base \cite{cap09} provide a good description, while  absolute values from the other models are too high \cite{rau97} or too low \cite{egi09}.  
\begin{table}[b]
\centering
\caption{Level density $(a)$, backshift $(\Delta)$  and spin cutoff ($\sigma$) parameters of the BSFG model predictions for $^{96}$Mo shown in Fig.~\ref{fig:ld}.
}
\begin{tabular}{ccccc}
\hline
\hline
Ref. & $a$ & $\Delta$ & $\sigma$(11.5 MeV) & $\sigma$(15.5 MeV) \\
& (MeV$^{-1}$) & (MeV) & ($\hbar$) & ($\hbar$) \\ 
\hline
\cite{cap09} & 11.25 & 1.14 & 5.32 & 5.77 \\
\cite{rau97} & 12.45 & 1.48 &  5.01 & 5.45 \\
\cite{egi09} &  9.56  & 0.82 & 4.20 & 4.42 \\ 
\hline
\hline
\end{tabular}
\label{tab:ld}
\end{table}

In order to compare with the Oslo results, the $1^-$ LD is converted to a total LD using a spin distribution function                    
\begin{equation}
\label{eq:sco}
f(J) \simeq \frac{2J+1}{2\sigma^2}\exp
	\left(-\frac{(J+\frac{1}{2})^2}{2\sigma^2}\right),
\end{equation}	
where $\sigma$ denotes the spin cutoff parameter.
Note that slightly different definitions of $f(J)$ are used in Refs.~\cite{cap09,rau97,egi09}. 
Values of $\sigma$ for the experimental energy range using the respective definitions are given in Tab.~\ref{tab:ld}.   
The model dependence of the conversion to toal LD is taken into account by averaging over the results from the three BSFG parameter sets and taking their variance as a measure of the model uncertainty. 
The resulting LD (red diamonds) is presented in Fig.~\ref{fig:ld} together with the Oslo results at lower excitation energies (blue squares) and  $s$-wave neutron capture (black circle) \cite{uts13}. 
The BSFG models are normalized to the value at $S_n$.
In particular, the RIPL-3 parameters \cite{cap09} provide a good description of all data over a large excitation energy range, consistent with a similar analysis for $^{208}$Pb \cite{bas16}.   

{\em Conclusions.}--
A new approach to test the Brink-Axel hypothesis is presented based on a study of the $(\vec{p},\vec{p}^\prime)$ reaction at 295 MeV and extreme forward angles. 
The extracted gamma strength function for the test case, $^{96}$Mo, agrees with results of compound nucleus $\gamma$ decay experiments \cite{gut05,lar10} indicating that the BA hypothesis holds in the energy region of the PDR, in contrast to results from the $(\gamma,\gamma^\prime)$ reaction \cite{rus09} and the claims of Ref.~\cite{ang12}.
This is an important finding since the BA hypothesis constitutes a general presupposition for astrophysical reaction network calculations.
The high energy resolution and selectivity of the data permits an extraction of the LD at excitation energies above the neutron threshold hardly accessible by other means.  
A consistent description of the LD with those of the $\gamma$ decay experiments can be achieved within BSFG models providing independent confirmation of the methods used to separate GSF and LD in Oslo-type experiments.       

While the present results support a use of the BA hypothesis for statistical model calculations of reaction cross sections in finite temperature environments, a general statement requires a systematic comparison of GSFs derived from $\gamma$ absorption and emission experiments in the energy range of the PDR over a broad range of nuclei.
For example, the role of deformation needs to be explored by comparing spherical and well-deformed cases with the present results for the moderatly deformed $^{96}$Mo.
Work along these lines is under way.        

We are indebted to the RCNP for providing excellent beams.
Discussions with E.~Grosse,  M.~Guttormsen, A.~C.~Larsen, R.~Schwengner, and S.~Siem are gratefully acknowledged.  
This work has been supported by the DFG under contract SFB 1245 and by MEXT KAKENHI Grant Number JP25105509.
C.A.B.\ acknowledges support by the U.S.\ DOE grant DE-FG02-08ER41533 and the U.S.\ NSF Grant No.\ 1415656.

\end{document}